\documentclass[12pt,preprint]{aastex}

 \shorttitle{On the Origin of
the Lightest Molybdenum Isotopes} \shortauthors{J.L. Fisker et al.}
\begin{document} \bibliographystyle{apj} \title{On the Origin of the
Lightest Molybdenum Isotopes} \author{Jacob Lund Fisker, Robert
D. Hoffman, and Jason Pruet} \affil{Lawrence Livermore National
Laboratory, 7000 East Avenue, P.O. Box 808, L-414, Livermore, CA
94551} \email{fisker1@llnl.gov,hoffman21@llnl.gov,pruet1@llnl.gov}
\begin{abstract} We discuss implications of recent precision
measurements for the ${}^{93}$Rh proton separation energy for the
production of the lightest molybdenum isotopes in proton-rich type II
supernova ejecta.  It has recently been shown that a novel
neutrino-induced process makes these ejecta a promising site for the
production of the light molybdenum isotopes and other 
``p-nuclei'' with atomic mass near 100. The origin of these isotopes
has long been uncertain.  A distinguishing feature of nucleosynthesis
in neutrino-irradiated outflows is that the relative production of
${}^{92}$Mo and ${}^{94}$Mo is set by a competition governed by the
proton separation energy of ${}^{93}$Rh. We use detailed nuclear
network calculations and the recent experimental results for this
proton separation energy to place constraints on the outflow
characteristics that produce the lightest molybdenum isotopes in
their solar proportions. It is found that for the conditions
calculated in recent two-dimensional supernova simulations, and also for a large
range of outflow characteristics around these conditions, the solar
ratio of ${}^{92}$Mo to ${}^{94}$Mo cannot be achieved.  This suggests
that either proton-rich winds from type II supernova do not
exclusively produce both isotopes, or that these winds are
qualitatively different than calculated in today's supernova models.

\end{abstract}


\cite{Burbidge57} described the synthesis of elements heavier than the
iron group in terms of three main processes: slow ($s$-process)
neutron capture, rapid ($r$-process) neutron capture, and proton
capture (the $p$-process). It was believed that proton capture could
account for the nuclei blocked from synthesis in neutron
processes by stable isotopes that prevent their production through
$\beta$-decay. Subsequent studies of
nucleosynthesis in stellar environments found the densities and
temperatures needed for the $p$-process difficult to obtain.  Today
the origin of the $p$-nuclei between $A=92$--$126$, and in
particular $^{92}$Mo and $^{94}$Mo, is one of the great outstanding
mysteries in nuclear astrophysics.

A number of potentially promising production sites and alternative
nucleosynthesis paths for producing the $p$-nuclei have been
proposed. These include neutron-rich outflows from nascent neutron
stars \citep{Hoffman96}, outflows from black hole accretion disks
\citep{Surman06}, and incredibly neutrino-irradiated neutron-rich
outflows from neutron stars \citep{Fuller95}. 
Recently it has been shown that a novel nucleosynthesis
process occurring in proton-rich supernova ejecta could efficiently
synthesize p-nuclei between strontium and palladium
\citep{Froehlich06b, Pruet06}.  In this process neutrino capture on
free protons produces a small reserve of free neutrons. These neutrons
induce (n,p) reactions that push the nuclear flow beyond
the waiting point nuclei with long weak decay half-lives (starting
with ${}^{64}$Ge). A feature of the neutrino-accelerated synthesis is
that all of the light p-process nuclei are made through decay of
radioactive proton-rich progenitors.

In this Letter we develop quantitative arguments to test and place
limits on neutrino-irradiated supernova outflows as the site
responsible for producing the two lightest molybdenum isotopes. The
observed ratio of these isotopes in the sun provides a key
diagnostic. We will show that this ratio in supernova outflows is
exquisitely sensitive to the proton separation energy for
$^{93}$Rh. As a result the production of the two lightest molybdenum
isotope in proton-rich winds with a relative abundance equal to that
observed in the sun is only possible for a narrow range of proton
separation energies near $S_p = 1.65$ MeV.  This value was consistent
with the previous empirically-based estimate $S_p=2.0\pm0.5$\,MeV
\cite{Audi03b}. However, recent precision measurements by
\cite{Fallis08} and \cite{Weber08} find that $S_p = 2.00 \pm 0.01 {\,
\rm MeV}$. Because these measurements do not give agreement with the
diagnostic afforded by the solar ratio, there is an indication that
proton-rich supernova outflows are not exclusively responsible for
producing both light molybdenum isotopes.

\begin{figure}[t]
\includegraphics[angle=270,width=\columnwidth]{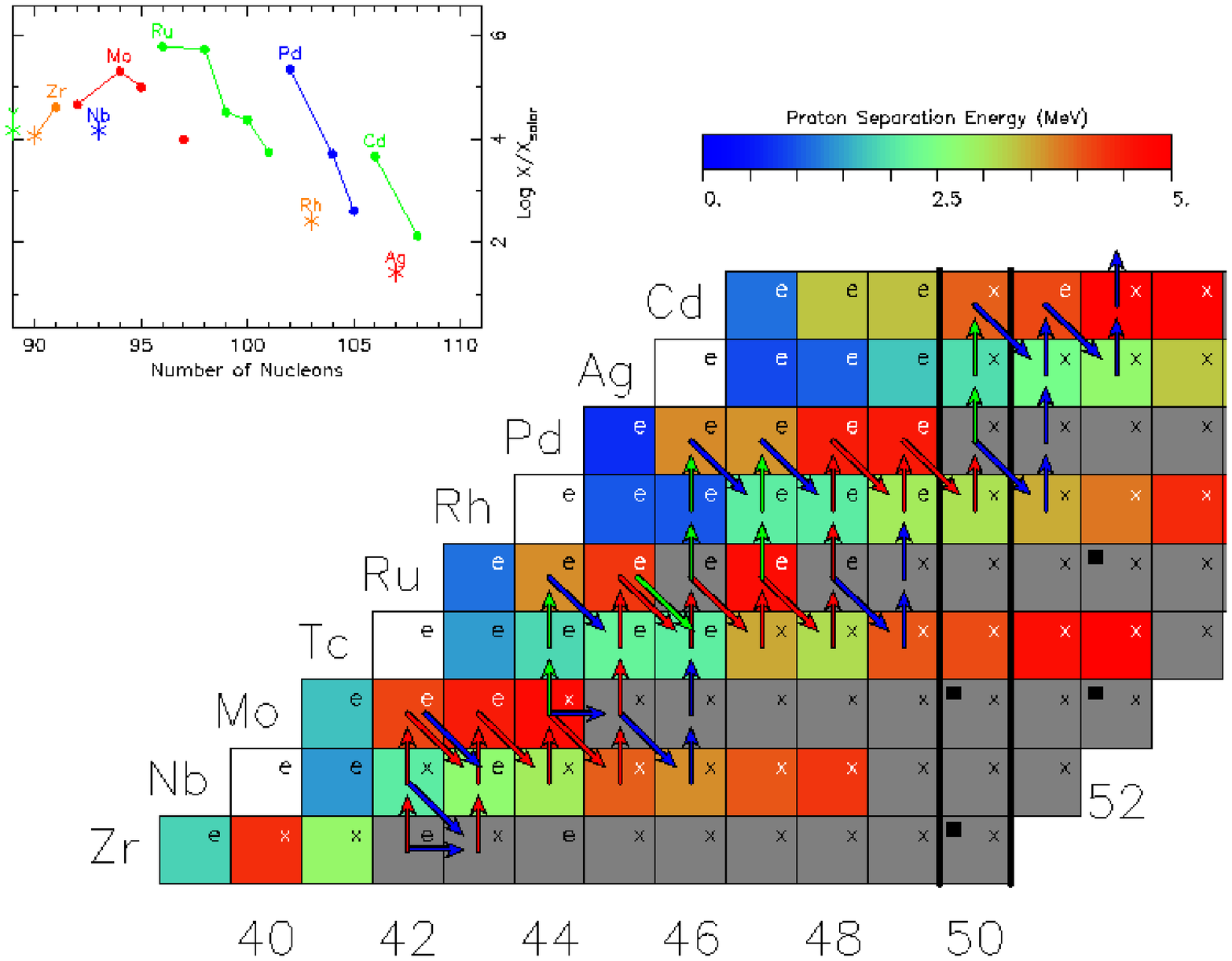}
\caption{
Net nuclear flows important in setting the abundance of
$^{92}$Mo relative to $^{94}$Mo. 
At the time shown ${\rm T}_9 = 2.06, \rho_5 = 2.74, {\rm and \ } Y_e = 0.561$.
Each isotope is color coded to the value of its proton separation energy
(red is 5 MeV, blue is 0 MeV, gray and white are above and below this range
respectively).
An "x" indicates the isotope has an experimentally
measured mass excess in the most recent mass evaluation \cite{Audi03b},
an "e" represents their extrapolation from measured masses.
The arrows indicate the dominant net nuclear
flows with color representing strength. All net flows within
a factor of 5 of the largest flow in this figure
are colored red, those between 5 and 10 are green, and between 10 and 50 are blue. The inset shows the production of the light $p$-nuclei relative to the solar abundances. The most abundant isotope in the sun is shown as an asterisk.
\label{netflowfig_sp}}
\end{figure}

\begin{figure}[t]
{\includegraphics[angle=270,width=10cm]{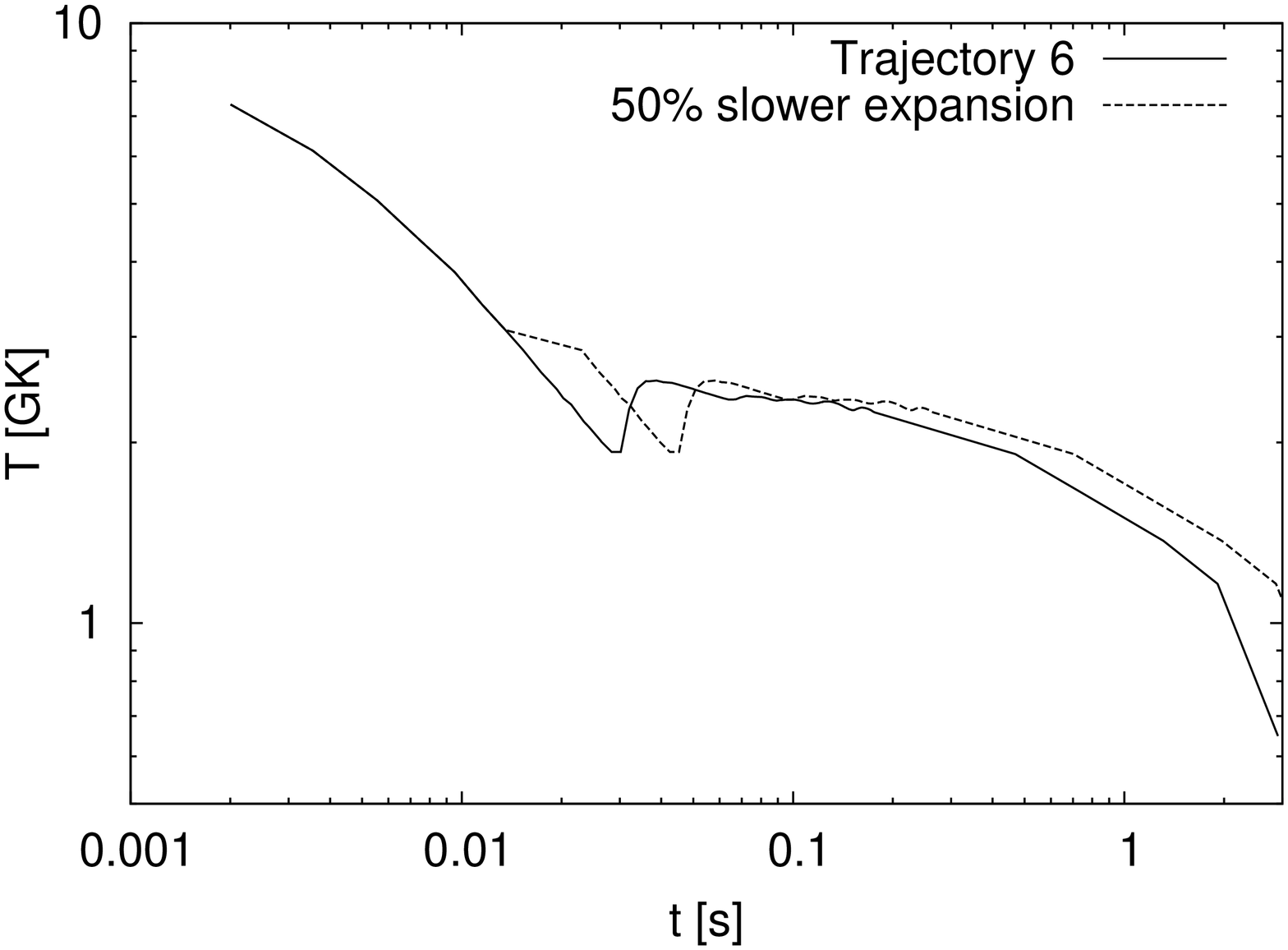}}
\vfill
{\includegraphics[angle=270,width=10cm]{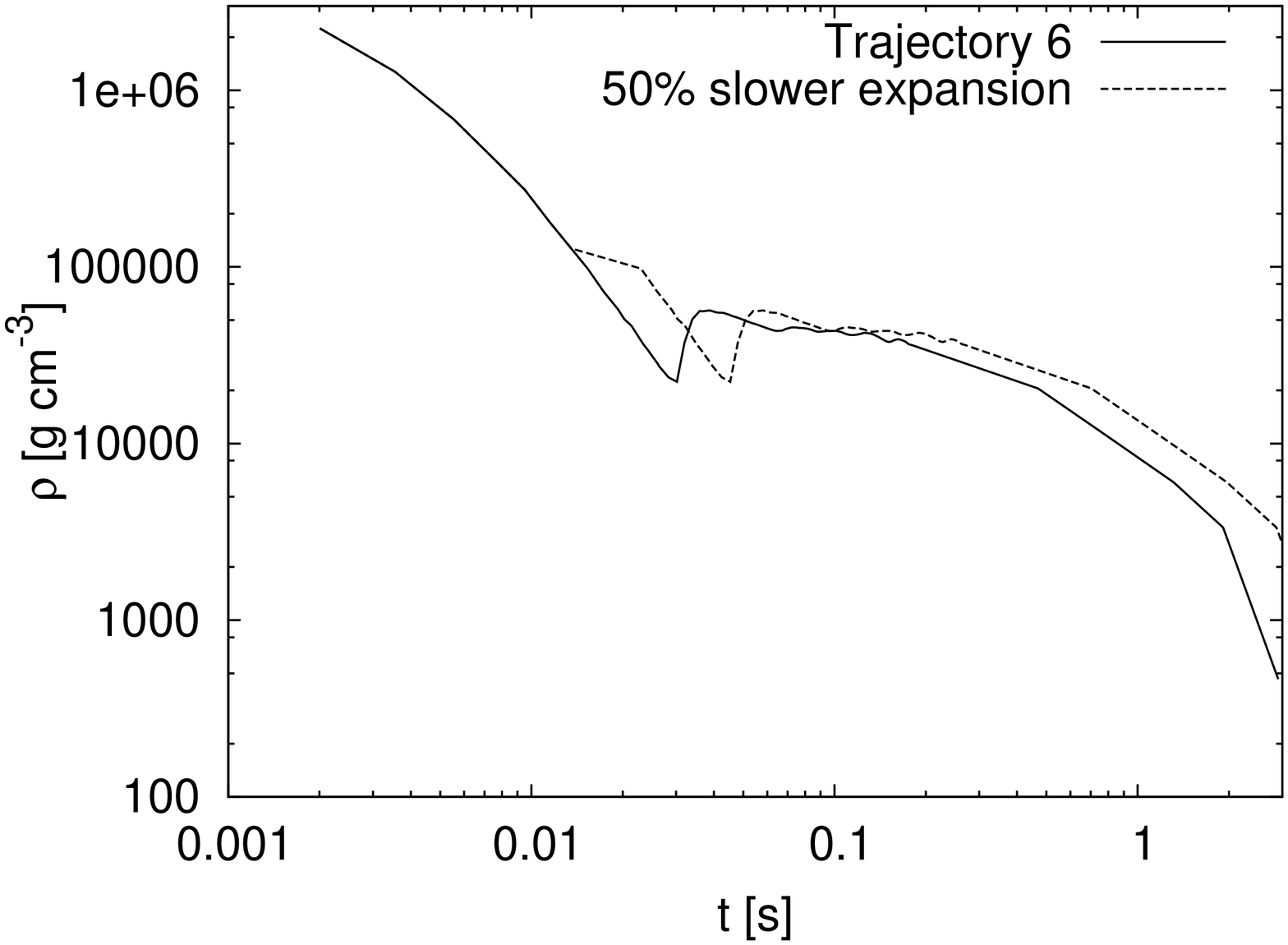}}
\vfill
{\includegraphics[angle=270,width=10cm]{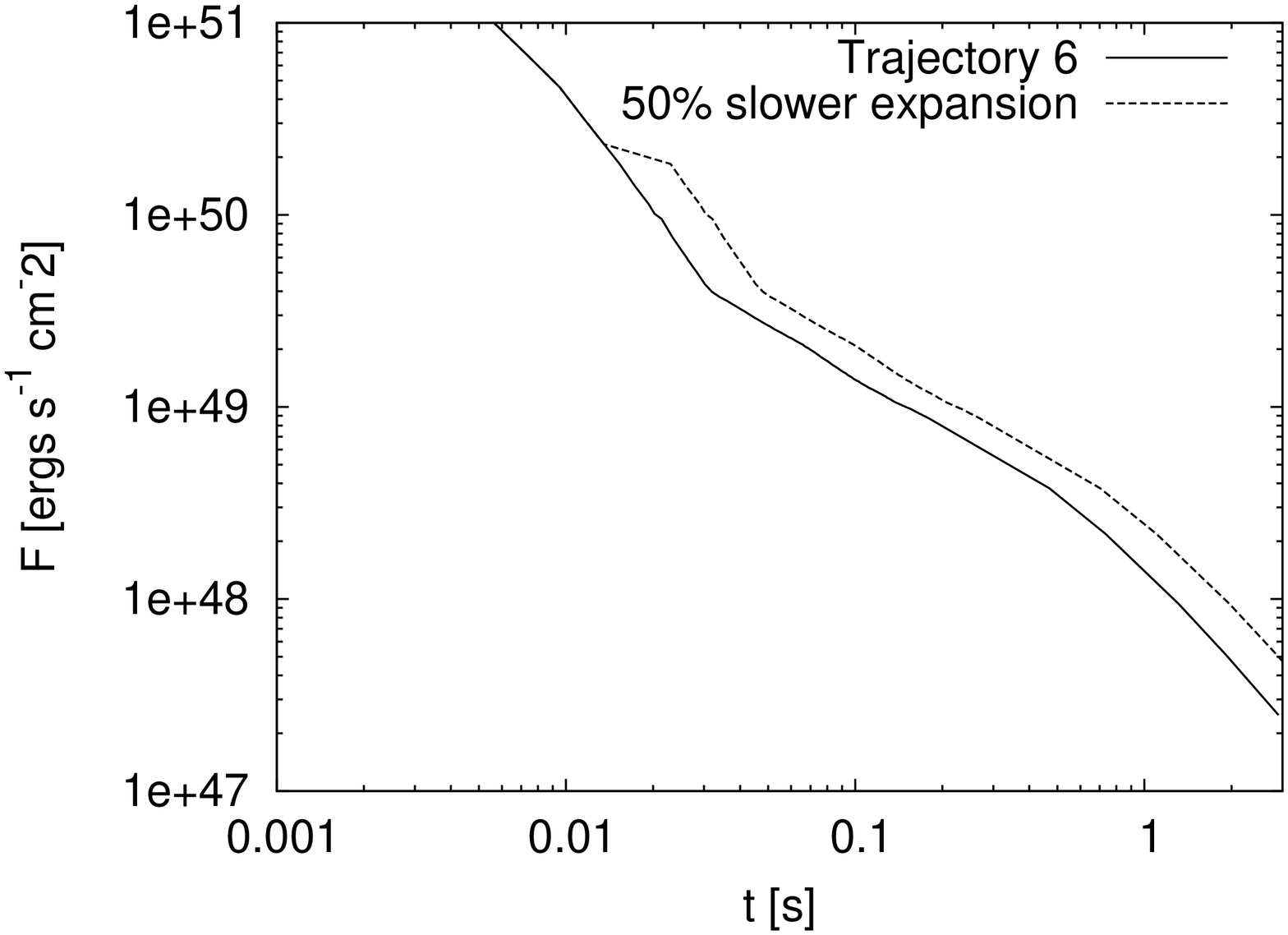}}
\caption{Shown are the evolution of the temperature, density, and total neutrino flux for the baseline ``trajectory 6'' as well as the evolution when the dynamic time scale is changed by 50\%.}\label{baseline}
\end{figure}

To set the stage for a quantitative discussion we show in Figure
\ref{netflowfig_sp} the net nuclear flows governing synthesis of
the lightest molybdenum isotopes in a supernova outflow. This outflow
is taken from the 2D supernova simulations of \cite{Janka03} and
corresponds to "trajectory 6" in the nucleosynthesis study of
\cite{Pruet06}. This is one of only two outflow trajectories
calculated by \cite{Janka03} that efficiently synthesized A$>$90
p-nuclei. We will take this outflow trajectory, which is characterized
by an entropy per baryon $s/k_B=77$ and electron fraction $Y_e=0.57$,
as our baseline. Figure \ref{baseline} shows conditions during the
first three seconds post core bounce in trajectory 6.

At temperatures larger than about 1 billion degrees, nuclear flows
governing synthesis of the light molybdenum isotopes are regulated by
a competition between $(p,\!\gamma)$ and the inverse $(\gamma,p)$
reactions. This balance is set chiefly by proton separation
energies. From Figure \ref{netflowfig_sp} it can be seen that the
proton separation energies for Rh isotopes with mass numbers between
89 and 91 are small. As a result, the even-even nucleus $^{92}$Pd is
not appreciably populated and does not serve as a progenitor for
$^{92}$Mo. Instead the main flow path detours along Z=44 (Ru) until
reaching the N=48 isotonic line. It is these N=48 nuclei that serve as
progenitors for the light molybdenum isotopes.

\begin{figure}[t]
\includegraphics[angle=0,width=\columnwidth]{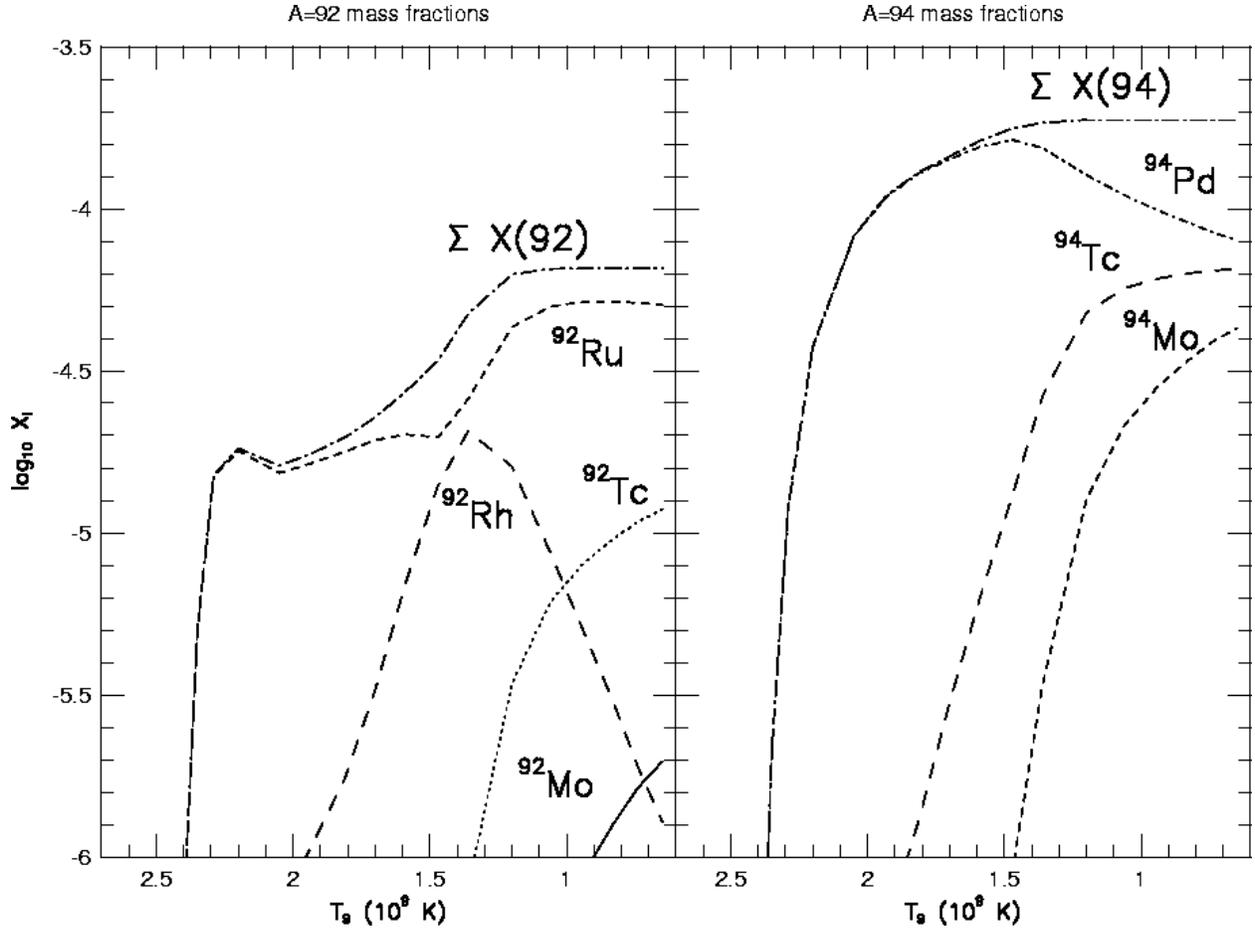}
\caption{Mass fractions of the $A=92$ and $A=94$ isobars affecting the final abundance
of $^{92}$Mo and $^{94}$Mo as a function of freeze out temperature. Note that most of the final yield
of these two molybdenum isotopes originates with $N=48$ isotones.
\label{combinedFig}}
\end{figure}

Figure \ref{combinedFig} depicts the late time evolution of the mass
fractions ($X$) for the $A=92$ and $94$ isobars that contribute to
$^{92}$Mo and $^{94}$Mo. Approximately $90\%$ of the final $^{92}$Mo
abundance results from $\beta^+$ decays starting at $^{92}$Ru. A
larger fraction of the final $^{94}$Mo inventory is attributed to
decays starting with $^{94}$Pd.

Since $N=48$ isotones are the main progenitors for the two light
molybdenum isotopes, the final abundance of $^{92}$Mo relative to
$^{94}$Mo is set principally by nuclear flow out of $^{92}$Ru.  Proton
capture on this nucleus produces $^{93}$Rh, which in turn leads to
efficient synthesis of the tightly bound and even-even
$^{94}$Pd. Though an important part of the nucleosynthesis occurs
after rates have become too slow to maintain nuclear statistical
equilibrium, we can gain some qualitative insights into the importance
of the proton separation energies by considering the case where the
outflowing ejecta is still hot.  Under this condition $^{92}$Ru and
its proton capture daughter $^{93}$Rh are in equilibrium with each
other. The relative abundance of the two nuclei is given by
\begin{equation}\label{eq:saha} \frac{Y_{93}}{Y_{92}}=5.15\cdot
10^{-11} \left(\frac{G_{93}}{G_{92}}\right) \left(\frac{\rho
Y_p}{T_{9}^{3/2}}\right) \exp(S_p(93)/T_9), \end{equation} where $Y$
represents the number fraction ($Y_i = \rho N_A n_i$), $G$ represents
the nuclear partition function, $S_p(93)$ is the proton separation
energy for $^{93}$Rh and $\rho$ is the density in g/cm$^{3}$.  Because
the proton separation energy for $^{93}$Rh is relatively small
compared to that for $^{94}$Pd, it plays the key role in setting the
final abundance of $^{92}$Mo relative to $^{94}$Mo. Qualitatively,
eq.~\ref{eq:saha} says that an increase in the $^{93}$Rh proton
separation energy tends to decrease the $^{92}$Mo/$^{94}$Mo ratio.

Quantitative results for the molybdenum production can be gained from
detailed nuclear network calculations. Our goal is to see whether the
solar ratio \cite{Lodders03} \begin{equation} \label{solarRatio}
\frac{X({}^{92}\textrm{Mo})}{X({}^{94}\textrm{Mo})} = 1.57
\end{equation} can be synthesized in proton-rich supernova ejecta.
Using the recently measured ${}^{93}$Rh proton separation energy, the
outflow characteristics from the simulation of \cite{Janka03}, and the
estimates of nuclear reaction rates described in \cite{Pruet06} gives
a ratio of the two lightest molybdenum isotopes of
${X({}^{92}\textrm{Mo})}/{X({}^{94}\textrm{Mo})}=0.35$.
This is about a factor of five too
small compared to the solar ratio in eq.~\ref{solarRatio}.

If today's supernova models were perfect then we could infer
directly from the discrepancy between the calculated and solar values
that proton-rich supernova winds are not exclusively responsible for
making both light molybdenum isotopes. However, there are potentially
important uncertainties in our understanding of neutron star
winds. From the perspective of trying to calculate nucleosynthesis
there are a few key aspects of the outflow.  These include the entropy
of the wind, the electron mole number, $Y_e$, and the dynamic
timescale characterizing the evolution of the wind \citep{Qian96}.

\begin{figure}[t]
\includegraphics[angle=270,width=\columnwidth]{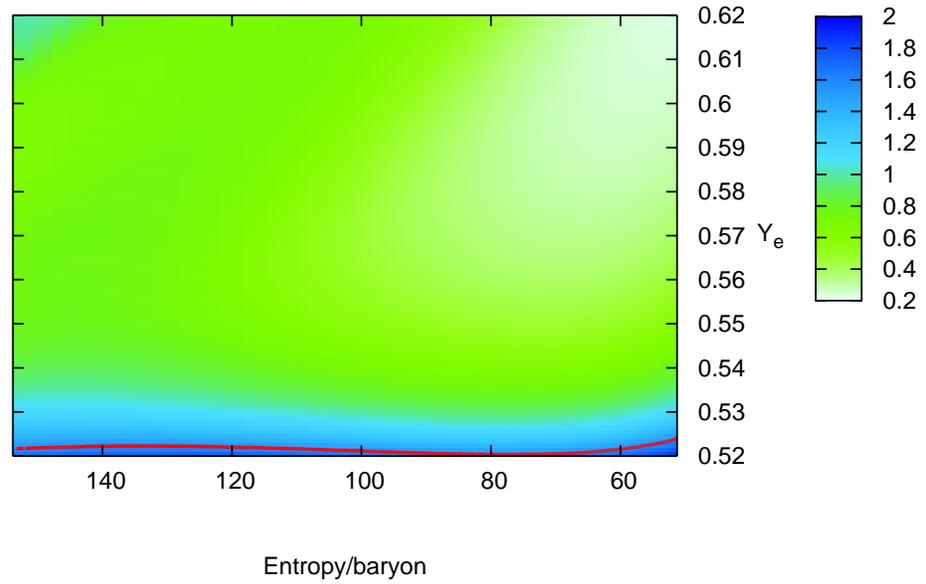}
\caption{The abundance ratio of $^{92}$Mo and $^{94}$Mo as a function of $Y_e$ and the entropy per baryon. The line very near $Y_e=0.52$ shows the solution where ${X({}^{92}\textrm{Mo})}/{X({}^{94}\textrm{Mo})} = 1.57$.}\label{enyefig}
\end{figure}

\begin{figure}[t]
\includegraphics[angle=270,width=\columnwidth]{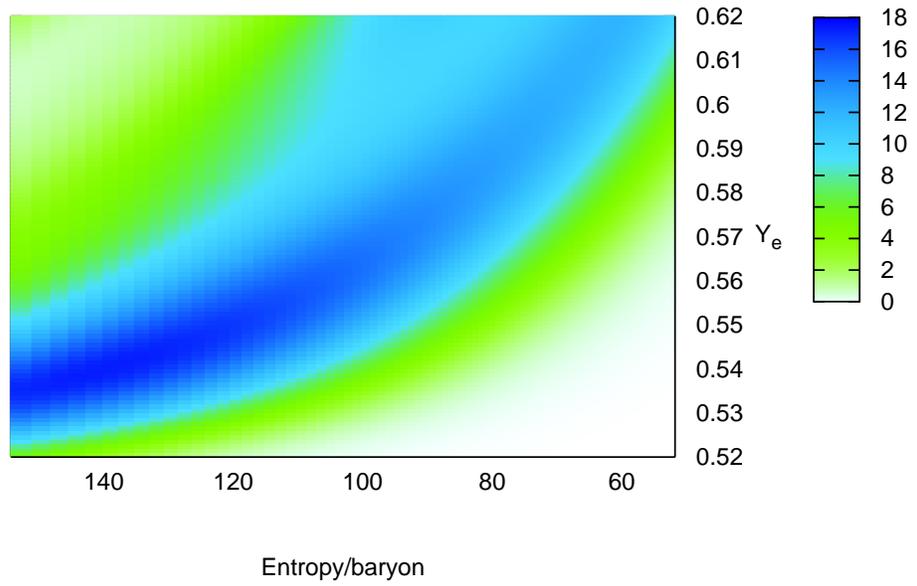}
\caption{The production factor for $^{92}$Mo (eq.~\ref{mo92productionf}) as a function of $Y_e$ and the entropy per baryon.}\label{enyemo92fig}
\end{figure}

To see if uncertainties in calculations of the outflow could account
for the discrepancy between the calculated and observed molybdenum
ratio, we calculated the nucleosynthesis for a great variety of assumptions
about conditions in the supernova. 
Results of modifying the electron fraction and entropy are shown in
Figure \ref{enyefig}. Entropy was varied by simply uniformly re-scaling the 
density as a function of time, which is approximately correct because of the 
relation $s/k_b\propto T^3/\rho$ valid in the regime important for nucleosynthesis \citep{Qian96}.
With the recently measured value for the proton
separation energy the solar ratio can only be achieved for $Y_e\approx
0.52$.  However, as Figure \ref{enyemo92fig} shows, at this electron
fraction the overall production of molybdenum plummets unless the
entropy is larger than about 120. This is much higher than the value
of 77 found in the supernova calculations. For smaller entropies the
production factor for $^{92}$Mo, which provides a measure of the total
amount of the isotope produced, is
\begin{equation}\label{mo92productionf} P(^{92}{\rm
Mo})=\left(\frac{X(^{92}{\rm Mo})}{X_{\odot}(^{92}{\rm
Mo})}\right)\left(\frac{M_{\rm wind}}{M_{\rm ejecta}}\right) <0.07,
\end{equation} where $M_{\rm wind}\approx 1.04\cdot10^{-3}\,M_{\odot}$
is the total mass of material producing $^{92}$Mo in the calculations
of \cite{Janka03} and $M_{\rm ejecta}\approx 13.5\,M_{\odot}$ is the
total mass of material ejected by the supernova. To account for the
solar abundance of isotopes attributed to type II supernovae, the
production factor has to be approximately 10 \citep{Timmes95}. This is found 
by calculating the production factor characterizing isotopes such as $^{16}$O
that are believed to be chiefly produced in supernovae\citep{Woosley95}. 
The needed production factor is also arrived at through detailed galactic
chemical evolution studies that describe stellar mass recycling and
galactic inflow and outflow \citep{mathews92,Timmes95}. 
A production factor near 10 
implies that if winds with $Y_e
\approx 0.52$ were responsible for the two light molybdenum isotopes
then the total mass of material producing these
isotopes would have to be more than an order of magnitude larger than
that calculated in current supernova models.

We also investigated the influence of uncertainties in the dynamic
timescale on the ratio of the light Mo isotopes. At high temperatures
the e--folding time for density sets the production of ``seed'' nuclei
with mass greater than A=12 and at low temperatures it sets the net
number of neutrino captures on protons. To approximately gauge the
influence of uncertainties in the dynamic timescale we simply scale
the time coordinate for our baseline trajectory by +50\% (see Figure
\ref{baseline}).  The ratio of the light Mo isotopes was essentially
unchanged.  Given this weak sensitivity it seems reasonable to neglect
errors associated with dynamic timescale. As well, the study of
\cite{Qian96} find that a factor of two change in dynamic timescale
corresponds to a factor of two change in neutron star radius or
neutrino luminosity.

In the same way that one can determine the outflow characteristics
that reproduce the observed molybdenum ratio, we can also determine
the ${}^{93}$Rh proton energy consistent with the observed ratio. For
entropies less than 140, and outflows that give a $^{92}$Mo production
factor larger than 0.7, we find that the solar ratio can only be
recovered if $S_P(93)=1.65\pm0.1$ MeV. This is ruled out by the recent
mass measurements.

It is also possible that uncertainties in other nuclear physics inputs
apart from the $^{93}$Rh proton separation energy could explain the
discrepancy between the calculated and solar molybdenum ratio. We
studied variations consistent with current experimental uncertainties
for the $^{91}$Rh proton separation energy and the $^{92}$Rh proton
separation energy. Uncertainties in these have effectively no impact
on the calculated Mo ratio. We also studied the impact of
uncertainties in the charged particle capture cross sections that
affect the production of $^{92}$Mo and $^{94}$Mo by increasing the
${}^{90}\textrm{Ru}(p,\gamma)^{91}\textrm{Rh}$,
${}^{91}\textrm{Ru}(p,\gamma)^{92}\textrm{Rh}$,
${}^{92}\textrm{Ru}(p,\gamma)^{93}\textrm{Rh}$, and
${}^{94}\textrm{Pd}(p,\gamma)^{95}\textrm{Ag}$ rates by a factor
100. Because these nuclei nearly are in equilibrium, these variations
result in a $<1\%$ change in the production factors. Though we can't
strictly rule out the possibility, it appears that nuclear physics
uncertainties cannot account for the discrepancy in the molybdenum
ratio.

In summary, our study of nucleosynthesis using the newly measured
value for the $^{93}$Rh proton separation energy suggests the
proton-rich supernova ejecta are not exclusively responsible for
producing the two lightest molybdenum isotopes unless conditions in
these ejecta are quite different than indicated by recent supernova
calculations.  In particular, the free proton fraction would have to
be about three times smaller, and the entropy about fifty units
higher. This would have fairly dramatic implications not just for
molybdenum production but also for the nucleosynthesis as a whole. It
may be interesting to note that the need for a decrease in electron
fraction and an increase in entropy also plagues calculations of the
$r$-process that might occur later in the evolution of the supernova
\citep{Qian96}.

\acknowledgments
We wish to thank S. E. Woosley and H.-T. Janka for very useful discussions and R. Buras for providing the supernova model trajectories used in this study. 
This work was performed under the auspices of the U.S. Department of Energy by Lawrence Livermore National Laboratory in part under Contract W-7405-Eng-48 and in part under Contract DE-AC52-07NA27344.


\clearpage

\end{document}